\documentclass[a4paper,11pt]{article}
\usepackage{pos}
\usepackage{amsmath}
\newcommand{\tmb}[1]{{\mbox{\tiny{#1}}}}

\title{Sampling Nambu-Goto theory using Normalizing Flows}

\author[a,b]{M. Caselle}
\author*[a,b]{E. Cellini}
\author[a,b]{A. Nada}

\affiliation[a]{Department of Physics,  University of Turin,\\
  Via Pietro Giuria 1, I-10125 Turin, Italy}

\affiliation[b]{INFN, Turin,\\
Via Pietro Giuria 1, I-10125 Turin, Italy}

\emailAdd{elia.cellini@unito.it}

\abstract{
Effective String Theory (EST) is a non-perturbative framework used to describe confinement in Yang-Mills theory through the modeling of the interquark potential in terms of vibrating strings. An efficient numerical method to simulate such theories where analytical studies are challenging is still lacking. However, in recent years a new class of deep generative models called Normalizing Flows (NFs) has been proposed to sample lattice field theories more efficiently than traditional Monte Carlo methods. In this contribution, we show a proof of concept of the application of NFs to EST regularized on the lattice. Namely, we introduce Physics-Informed Stochastic Normalizing Flows and we use them to sample the Nambu-Goto string action with two goals: use the known analytical results of this theory as a benchmark and demonstrate the efficiency of our method in obtaining new results of physical interest and in particular in providing a numerical proof for a conjecture regarding the width of the string.
}

\FullConference{The 40th International Symposium on Lattice Field Theory (Lattice 2023)\\
July 31st - August 4th, 2023\\
Fermi National Accelerator Laboratory\\}
\begin{document}
\maketitle

\section{Introduction}
In a recent work~\cite{caselle:2023s}, a new numerical method based on deep generative algorithms has been provided to sample efficiently configurations of an Effective String Theory (EST) regularized on the lattice. EST is a non-perturbative tool used to study confinement in Yang-Mills theories in which the chromoelectric flux tube that connects a quark and an antiquark is modelled in terms of a thin vibrating string. Specifically, in EST the correlator between Polyakov loops is related to the partition functions of the EST. The simplest choice for the EST action is the Nambu-Goto (NG) string. It is anomalous at quantum level; however, in the large-distance limit, the anomalies vanish and several analytical studies of the NG partition function together with lattice calculation of pure gauge theories proved that EST is an highly predictive effective model \cite{Aharony:2013ipa,Caselle:2021eir}. 
Although the NG partition function is well known, much less is known for the correlation functions. In principle, the NG action can be regularized on the lattice and treated as a spin model. However, due to the non-linearity of the action, standard algorithms are not efficient at sampling such a model. Thus, the methods presented in \cite{caselle:2023s} relaying on Normalizing Flows \cite{rezende2015variational}, a class of deep generative models able to sample from Lattice Field Theory \cite{Albergo1,Nicoli2021}, represent a promising approach to numerical calculations of EST. Nevertheless, the Continuous Normalizing Flows (CNFs) \cite{Gerdes:2022eve} used in \cite{caselle:2023s} suffer from poor scaling. Therefore, in this contribution, we present a Physics-Informed Stochastic Normalizing Flow \cite{wu2020stochastic,Caselle:2022acb,Abbott:2022zsh} that provides an highly efficient sampler for the NG theory. 

In our studies, we focused the high temperature regime of the string, we used the well-known analytical solution of the partition function as a benchmark and tested the effectiveness of the PI-SNFs by providing a numerical proof of a conjectured non-perturbative solution of the width of the string in the NG theory \cite{Caselle:2010zs}, a correlation function that is related to the density of the chromoelectric flux tube.

\section{Lattice Nambu-Goto String}
In the case of a $d=2+1$ target Yang-Mills theory, the corresponding Nambu-Goto action regularized on a two-dimensional lattice using a "physical gauge" can be written as \cite{caselle:2023s}:
\begin{equation}\label{eq:NG}
    S_\tmb{NG}=\sigma \sum_{x \in \Lambda} \biggl(\sqrt{1+(\partial_{\mu}\phi(x))^2/\sigma}-1\biggr) 
\end{equation}
where $\Lambda$ is a square lattice of size $L\times R$ with index $x=(x^0,x^1)$ representing the worldsheet of the string and lattice step $a=1$; $\phi(x) \in \mathbb{R}$ is a real scalar field representing the transverse degrees of freedom of the string and $\sigma$ is the string tension, the coupling of the theory. We fix periodic boundary conditions $\phi(x^0,x^1)=\phi(x^0+L,x^1)$ along the temporal extension of length $L$, and Dirichlet boundary conditions $\phi(x^0,0)=\phi(x^0,R)=0$ along the spatial extension of length $R$, which represent the Polyakov loops $P$ and $P^{\dagger}$. Following this choice, we can identify the physical temperature as the inverse of $L$. Thus, the high temperature region is the limit in which $R\gg L$; in this regime, the analytical solution for the partition function is:
\begin{equation}\label{eq:logZsolution}
-\log Z= (\sigma(L)-\sigma) RL+\frac{1}{2}\log (2R/L)+d.t.
\end{equation}
where $\sigma(L)=\sigma\sqrt{1-\frac{\pi}{3\sigma L^2}}$ encodes the dependence of the string tension on the physical temperature and $d.t.$ represents the divergent terms which arise in the finite-size lattice regularize string (see \cite{caselle:2023s}). The width of the regularized Nambu-Goto string can be computed as: $ \sigma w^2(\sigma, L,R)=\langle \phi^2(x^0,R/2)\rangle_{x^0}$, where the expectation value $\langle ... \rangle_{x^0}$ is evaluated also over the temporal extension $x^0$ (because of the translation invariance imposed by the periodic boundary conditions). The conjectured non-perturbative solution state that \cite{Caselle:2010zs}:
\begin{equation}\label{eq:conjecture}
    w^2(\sigma, L,R)=\frac{1}{4}\frac{1}{\sigma(L)}\frac{R}{L}+...
\end{equation}

\section{Stochastic Normalizing Flows}
Stochastic Normalizing Flows (SNFs) \cite{wu2020stochastic,Caselle:2022acb} are a class of deep generative models that combines Normalizing Flows (NFs) with non-equilibrium lattice calculations based on the Jarzynski's equality \cite{Jarzynski1997,neal2001annealed,Caselle:2016wsw,CaselleSU3,Francesconi:2020fgi,Bulgarelli:2023ofi}. Jarzynski's equality \cite{Jarzynski1997} is an identity in non-equilibrium statistical mechanics which states that ratios of partition functions between two equilibrium states $\eta_{in}$ and $\eta_{fin}$ can be computed as an average of out-of-equilibrium processes: $\frac{Z_{\eta_{fin}}}{Z_{\eta_{in}}}=\langle \exp(-w(\phi_0,\phi_1,...,\phi_N))\rangle_f $, where $\eta(t)$ is a protocol that drives out of equilibrium the initial state $\eta_{in}=\eta(t_{in})$ toward the target $\eta_{fin}=\eta(t_{fin})$. $w(\phi_0,\phi_1,...,\phi_N)$ is the dimensionless work done on the system during the transformations and the average $\langle ... \rangle_f$ is taken over all the possible path connecting, in the phase space, the states $\eta_{in}$ and $\eta_{fin}$. An intriguing feature of the Jarzynski's equality is that only the initial configurations must be at thermodynamic equilibrium while all the others configurations, included the target $\eta_{fin}$, can be arbitrarily far from the equilibrium. Jarzynski's equality can be exploited to build a Markov Chain Monte Carlo algorithm that we denote as "stochastic evolutions" (equivalent to Annealing Important Sampling~\cite{neal2001annealed}) able to compute partition functions ratios and expectation values. Given an initial "prior" distribution $q_{\eta_{0}}=e^{-S_{\eta_{0}}[\phi_0]}/Z_{\eta_{0}}$ with samples $\phi_0 \in \mathbb{R}^d$, and the target $q_{\eta_{N}}=e^{-S_{\eta_{N}}[\phi_N]}/Z_{\eta_{N}}$, $\phi_N \in \mathbb{R}^d$, each stochastic evolution is made by a sequence of Monte Carlo updates with transition probabilities, defined in term of a Boltzmann distribution, which interpolate between the prior and the target following a discrete protocol $\eta$ that can be a set of couplings of the action $S$: at the $n-$step the distribution is $q_{\eta_{n}}[\phi]=\exp(-S_{\eta_n}[\phi])/Z_{\eta_n}$. In this framework, the dimensionless work can be computed as: $w(\phi_0,\phi_1,...,\phi_N)  = S_{\eta_N}[\phi_N]-S_{\eta_0}[\phi_0]-Q_h(\phi_0,\phi_1,....,\phi_N)$ with $Q_h(\phi_0,\phi_1,....,\phi_N) = \sum_{n=0}^{N-1}\{ S_{\eta_{n+1}}[\phi_{n+1}]-S_{\eta_{n+1}}[\phi_{n}] \}$.

 Stochastic evolutions can be combined with Normalizing Flows \cite{rezende2015variational}, to obtain SNFs. A NF is a parametric diffeomorphism $g_\theta: \mathbb{R}^d\to\mathbb{R}^d$ which maps the prior $q_0$ into an inferred distribution $q_\theta$ which approximates the target $p(\phi)=\exp(-S[\phi])/Z$. In general, the flow is built as a sequence of transformations $g_\theta=g_{\theta_N}\circ ...\circ g_{\theta_i}\circ ... \circ g_{\theta_1}$ implemented using neural networks with parameters $\theta_i$. A fundamental characteristic of NFs is that the density of the samples after the layer $g_{\theta_i}$ can be computed as: $q_{\theta_i}(g_{\theta_i}\circ ... \circ g_{\theta_1}(\phi_0)) =q_0(\phi_0)e^{-Q_{g_{\theta_i}\circ ... \circ g_{\theta_1}}}$ with $Q_{g_{\theta_i}\circ ... \circ g_{\theta_1}} =\sum_{n=0}^{i}\ln |\det J_n(\phi_n)|$,
 where $|\det J_n(\phi_n)|$ is the determinant of the Jacobian of the n-\textit{th} layer. SNFs arise by interleaving stochastic evolutions Monte Carlo updates and NFs layers $g_{\theta_i}$. The work is now defined as: $w(\phi_0,\phi_1,...,\phi_N)  = S_{\eta_N}[\phi_N]-S_{\eta_0}[\phi_0]-Q(\phi_0,\phi_1,....,\phi_N)$ where $Q$ is computed as the sum of the heat $Q_h$ of the stochastic updates and the sum of the logarithms of the determinants of the flows layers $Q_g$. The training of SNFs is done by minimizing the Kullback-Leibler divergence: $D_{KL}(q_0P_f||q_NP_r)=\langle w(\phi_0,\phi_1,...,\phi_N) \rangle_f + \ln{\frac{Z_N}{Z_0}}$, where $P_f$ and $P_r$ represent respectively the forward and the reverse probability of the non-equilibrium trajectories of the SNFs~\cite{Caselle:2022acb}. When the models are trained, using the Jarzynski's equality it is possible to compute partition functions: $Z_{N} =Z_{0}\langle \exp(-w(\phi_0,\phi_1,...,\phi_N))\rangle_f$, and general observables: $\langle \mathcal{O} \rangle_f =\langle \mathcal{O}(\phi_N) \exp(-w(\phi_0,\phi_1,...,\phi_N))\rangle_f/Z_N$.
 
As proposed in \cite{Abbott:2022zsh}, building algorithms including the physical a priori knowledge of the studied system leads to better models. In our work, we took inspiration from the $T\bar T$ irrelevant integrable perturbations \cite{Cavaglia:2016oda,Caselle:2013dra}, a class of inverse renormalization group, to build our Physics-Informed Stochastic Normalizing Flows (PI-SNF). Specifically, we used as a prior a massless free field regularized on the lattice: $S_{\eta_{0}}[\phi] =\frac{1}{2}\sum_{x\in \Lambda}(\partial_\mu\phi(x))^2$, and we drove out of the equilibrium these samples following the $T\bar T$ flux using the Nambu-Goto action itself as a protocol for the Monte Carlo updates: $S_{\eta_{n}}[\phi] =\sigma_n \sum_{x \in \Lambda} \biggl(\sqrt{1+(\partial_{\mu}\phi(x))^2/\sigma_n}-1\biggr)$ with $\sigma_{n-1} > \sigma_{n}$.

\section{Results}\label{sec:results}
As mentioned in the introduction, the main focus of our numerical study is provide a highly efficient algorithm for EST and prove numerically the non-perturbative solution of the width of the Nambu-Goto string in the high temperature regime which is inaccessible with the CNFs used in~\cite{caselle:2023s}. 

We used a PI-SNF with $202$ "layers", each one is made by composing one affine and one stochastic block. The affine block is a composition of two affine coupling layers~\cite{Dinh:2017}, with even-odd masks, that share the same convolutional neural network with three hidden layers with $3\times 3\times 16$ kernels and one $2$ channels output layer. The stochastic block is built using Hybrid Monte Carlo (HMC) updates with leapfrog integration with $10$ integration steps and $\epsilon_{HMC}=0.1$. In each HMC step, we fixed the NG action following a linear protocol in the inverse of the string tension $\sigma$. 

\begin{figure}
  \centering
  \includegraphics[scale=0.7,keepaspectratio=true]{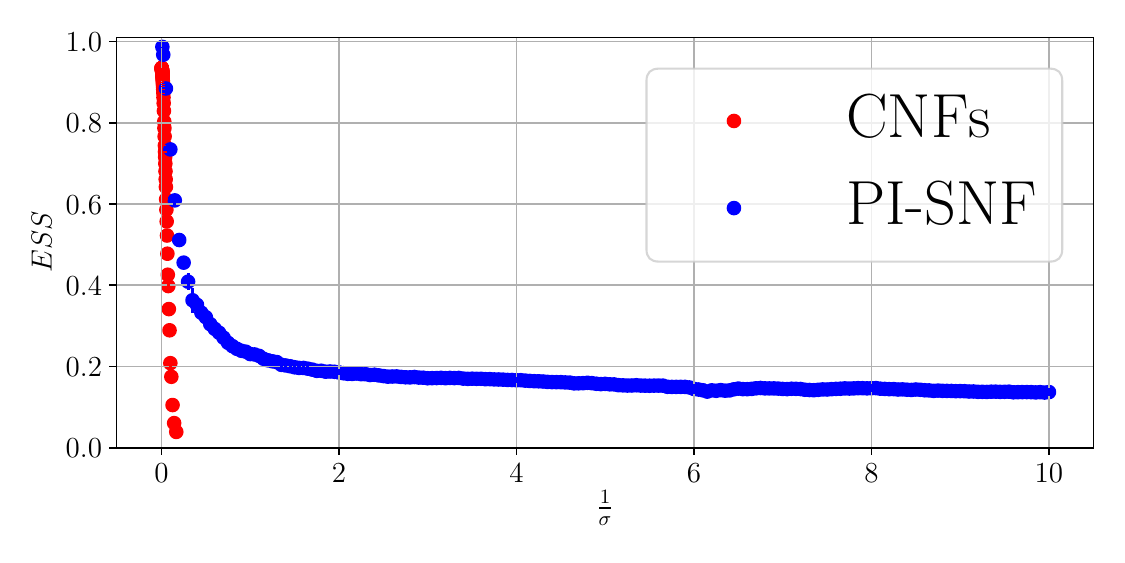}
  \caption{ESS as a function of $\frac{1}{\sigma}$ for the CNFs used in~\cite{caselle:2023s} and the PI-SNF described in this work.}
  \label{fig:scaling}
\end{figure}

A popular metric used to compare the performance of flow-based sampler is the Effective Sample Size: $\mbox{ESS} = \frac{\langle \Tilde{w} \rangle^2 }{\langle \Tilde{w}^2 \rangle}$. In fig.~(\ref{fig:scaling}), we report a comparison on $40\times 40$ lattices between the CNFs of~\cite{caselle:2023s} and the PI-SNF used in this contribution which clearly shows the outstanding performance of the stochastic flows compared to the continuous.

\begin{figure}
  \centering
  \includegraphics[scale=0.6,keepaspectratio=true]{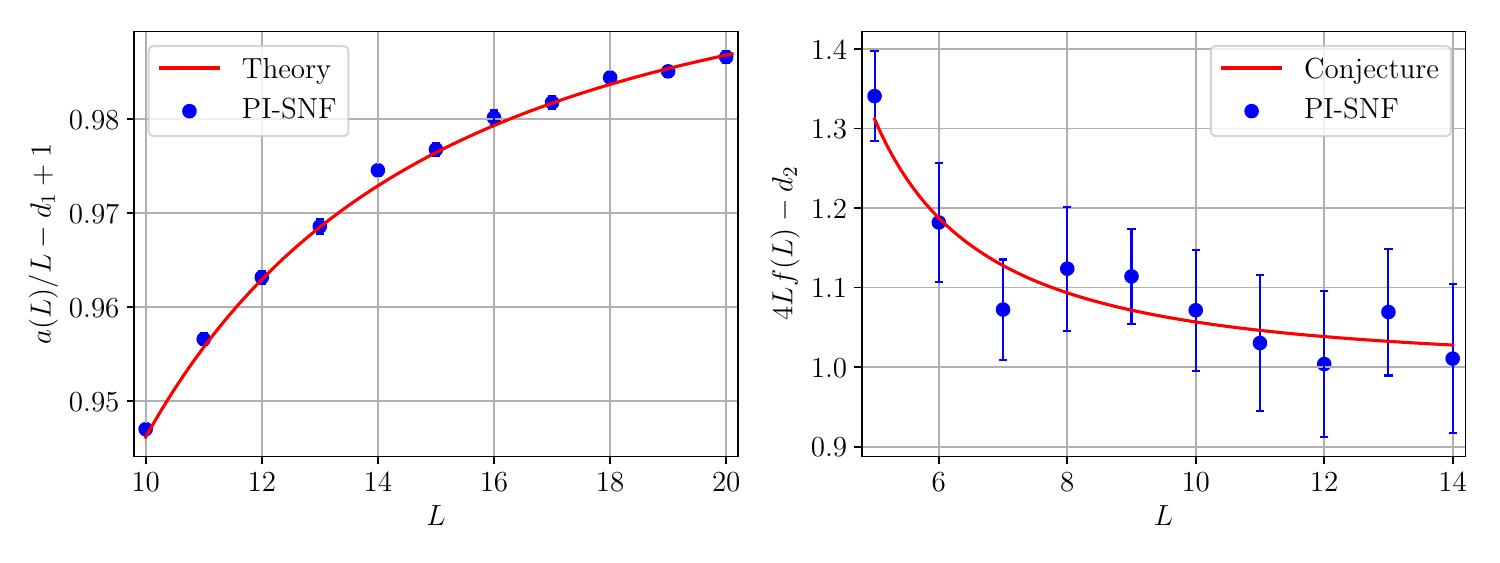}
  \caption{Results for $a(L)/L-d_1+1$ (left plot) and $4Lf(l)-d_2$ (right plot).}
  \label{fig:results}
\end{figure}

As a benchmark for our simulations we study the partition functions and we selected $286$ combinations of $L$ and $R$ with fixed $\sigma=0.1$ and average $\overline{\mbox{ESS}}$ equal to $0.249(6)$. We first fitted in $R$ the following expression:
\begin{equation}\label{eq:HTZR}
    -\log Z(L,R)=a(L)\sigma R+b(L)+c(L)\log(R),
\end{equation} 
then, fitting the coefficients $a(L)$ in $L$ according to:
\begin{equation}\label{eq:HTZL}
    a(L)= L\biggl(\sqrt{1-k_1\frac{1}{\sigma L^2}}+d_1-1\biggr)
\end{equation}
leads to a value for $k_1=1.03(1)$ which is in agreement (within $2$ times the standard deviation) with the theoretical prediction $\pi/3=1.047...$. The results of the fit are listed in table~\ref{tab:logZ}; in fig.~(\ref{fig:results}) the physical term $a(L)/L-d_1+1$ is compared to the analytical solution.  

One of the main goals of this contribution is the numerical proof of a conjecture about the non-perturbative solution of the Nambu-Goto string width in the high temperature regime. In this study, we considered $314$ combinations of $L$ and $R$ at fixed $\sigma=0.1$ and $\overline{\mbox{ESS}}=0.369(7)$. We first fitted in $R$ the expression:
\begin{equation}\label{eq:width1}
    \sigma w^2(L,R)=f(L)R+g(L)
\end{equation}
then in $L$:
\begin{equation}\label{eq:width2}
    f(L)=\frac{1}{4L}\biggl(\frac{1}{\sqrt{1-k_2\frac{1}{\sigma L^2}}}+d_2\biggr)
\end{equation}
where $d_2$ is a divergent term. We found an excellent agreement between $k_2=1.09(8)$ and the conjectured coefficient $\pi/3=1.047...$ (results listed in table~(\ref{tab:width})). As for the partition function, in fig.~(\ref{fig:results}) we plotted the physical term $4Lf(l)-d_2$ and compared it to the conjectured solution.
\begin{table}
\parbox{.45\linewidth}{
\centering
\begin{tabular}{|c c c c c|} 
 \hline
 $L$ & $a(L)$ & $b(L)$ & $c(L)$ &  $\chi^2/d.o.f.$\\ [0.5ex]
 \hline\hline
10 &  -123.271(4) & 15.26(9) & 0.54(3) & 0.47  \\
\hline
11 &  -135.493(6) & 16.9(1) & 0.52(4)  & 0.95 \\
\hline
12 &  -147.732(7) & 18.3(2) & 0.55(5)   & 1.10 \\
\hline
13 &  -159.972(9) & 20.0(2) & 0.54(6) & 1.57\\
\hline
14 &  -172.194(6) & 22.0(1) &  0.39(1)  & 0.82 \\
\hline
15 &  -184.460(8) & 23.2(2) & 0.50(6)  & 1.04 \\
\hline
16 &  -196.70(1) & 25.0(2) & 0.44(7)  & 1.67 \\
\hline
17 &  -208.97(1) & 26.4(2) & 0.52(7)  & 1.43 \\ 
\hline
18 &  -221.215(7) & 28.2(2) & 0.43(5)  & 0.66 \\ 
\hline
19 &  -233.492(7) & 29.4(2) & 0.56(5)  & 0.58 \\ 
\hline
20 &  -245.75(1) & 31.0(2) & 0.55(7)  & 1.17 \\ 
 \hline
\end{tabular}
\\
\vspace{0.3cm}
\begin{tabular}{|c c c|} 
 \hline
 $k_1$ & $d_1$ & $\chi^2/d.o.f.$ \\ [0.5ex]
 \hline\hline
 $1.03(1)$ & $-12.2742(4)$ & $1.40$ \\
 \hline
\end{tabular}
\caption{Results for the coefficients of the fit of $\log Z$ of eq.~(\ref{eq:HTZR}) (upper table) and eq.~(\ref{eq:HTZL}) (lower table)}
\label{tab:logZ}
}
\hfill
\parbox{.45\linewidth}{
\centering
\begin{tabular}{|c c c c|} 
 \hline
 $L$ & $f(L)$ & $g(L)$ & $\chi^2/d.o.f.$\\ [0.5ex]
 \hline\hline
5 &  0.681(3) & 2.7(1) & 0.9  \\
\hline
6 &  0.561(3) & 3.3(2)  & 0.74 \\
\hline
7 &  0.477(2) & 3.8(1)   & 0.64 \\
\hline
8 &  0.419(2) & 3.8(1) & 1.02\\
\hline
9 &  0.372(2) & 4.15(9)  & 0.60 \\
\hline
10 &  0.334(8) & 4.4(1)  & 0.81 \\
\hline
11 &  0.303(2) & 4.6(1)  & 1.15 \\
\hline
12 &  0.277(2) & 4.9(1)  & 1.24 \\ 
\hline
13 &  0.257(2) & 4.99(9)  & 0.87 \\ 
\hline
14 &  0.237(2) & 5.22(9)  & 1.21 \\ 
 \hline
\end{tabular}
\\
\vspace{0.3cm}
\begin{tabular}{|c c c|} 
 \hline
 $k_2$ & $d_2$ & $\chi^2/d.o.f.$ \\ [0.5ex]
 \hline\hline
 $1.09(8)$ & $12.28(2)$ & $0.25$ \\
 \hline
\end{tabular}
\caption{Results for the coefficients of the fit of $\sigma w^2$ of eq.~(\ref{eq:width1}) (upper table) and eq.~(\ref{eq:width2}) (lower table).}
\label{tab:width}
}
\end{table}
\section{Conclusion}\label{sec:conclusion}
In this contribution, we showed that physics-informed design and out-of-equilibrium thermodynamics, when applied to a flow-based sampler, can effectively address the scaling issues of the CNFs used in~\cite{caselle:2023s}. On one hand, the stochastic blocks can drive the evolution of the deterministic flow layers, while on the other hand, the $T\bar T$ perturbation offers a meaningful non-equilibrium protocol for the NG theory. Furthermore, we found a good agreement with the analytical solution of the partition function and provided the first numerical proof of the conjectured non-perturbative solution for the width of the NG theory in the high-temperature regime. Remarkably, in this study flow-based samplers have been used to obtain state-of-the-art results in a framework which is not a toy model. Future developments of this work will include an investigation of the so-called "rigid string" for which only a few analytical results have been provided.
\acknowledgments
The numerical simulations were run on machines of the Consorzio Interuniversitario per il Calcolo Automatico dell'Italia Nord Orientale (CINECA). We acknowledge support from the SFT Scientific Initiative of INFN. This work was partially supported by the Simons Foundation grant 994300 (Simons Collaboration on Confinement and QCD Strings) and by the Prin 2022 grant 2022ZTPK4E.

\bibliographystyle{JHEP}
\bibliography{biblio.bib}

\end{document}